# Testing the validity of THz reflection spectra by dispersion relations

K.-E. Peiponen *, E. Gornov and Y. Svirko

Department of Physics, University of Joensuu, P. O. Box 111, FI –80101 Joensuu, Finland

Y. Ino and M. Kuwata-Gonokami

Department of Applied Physics, University of Tokyo, Bunkyo-ku, Tokyo 113-8656, Japan, and Solution Oriented Research for Science and Technology (JST)

V. Lucarini

Department of Mathematics and Computer Science, University of Camerino, 62032 Camerino (MC), Italy

* Corresponding author, e-mail:kai.peiponen@joensuu.fi

**Abstract**

Complex response function obtained in reflection spectroscopy at terahertz range is examined with algorithms based on dispersion relations for integer powers of complex reflection coefficient, which emerge as a powerful and yet uncommon tools in examining the consistency of the spectroscopic data. It is shown that these algorithms can be used in particular for checking the success of correction of the spectra by the methods of Vartiainen et al [1] and Lucarini et al [2] to remove the negative misplacement error in the terahertz time-domain spectroscopy.

# 1. Introduction

During the past years terahertz (THz) spectroscopy has emerged as a powerful tool in investigation of dielectrics, semiconductors and superconductors in the far infrared spectral range. Since at THz frequencies we can measure amplitude rather than intensity of the probe wave, the signal in THz spectroscopy carries potentially more information on the properties of the material in comparison with that in conventional optical spectroscopy, which is based on intensity measurements. This opens new horizons in various applications including medicine, security control and materials inspection. However, since many materials that are non-transparent in THz region, it is crucial for applications to establish THz measurements in reflection configuration. Unfortunately, in reflection geometry, the sample-misplacement problem still remains as a bottleneck in THZ time-domain spectroscopy (THz-TDS) of opaque media.

Specifically, measurement of the complex reflection coefficient in the THz spectral range involves normalization of the signal reflected from the studied sample to the reference signal (usually it is a signal obtained by reflecion of a THz wave from a metal plate). However, the phase of the signal in THz-TDS depends on the position of the reflected surface in the experimental setup. This implies that one has to place the sample and the reference in the same position. Otherwise due to the sample misplacement the measured reflection spectrum is not correct but distorted because of a finite phase shift of the THz wave reflected from the sample with respect to that of a reference. Recently we have presented two methods to correct the phase of the THz-TDS signal [1, 2]. The first one is based on maximum entropy method (MEM) [1], which relies purely on mathematical model in information theory, and it is to some extent complicated. The second method is based directly on singly subtractive Kramers-Kronig relations (SSKK) [2]. These two methods allow one to reveal the correct complex reflection coefficient of the medium from erroneous THz-TDS signal and emerge as powerful tools for both fundamental and applied THz studies, in particular, for materials inspection. Importantly, in addition to the correction of the spectrum MEM- and SSKK methods enable us to avoid fine-tuning in THz reflection experiments.

In this paper we demonstrate that negative misplacement error (when optical path in the setup with sample is shorter than that with the reference) in the band-limited THz-TDS data can be detected by using dispersion relations without data extrapolations beyond the measured spectral range. Moreover such dispersion relations can also be employed for checking how well the correction of the spectrum was succeeded. Here we make use of dispersion relations for the integer powers of the complex reflection coefficient given by Smith and Manogue (S&M) [3]. In addition, we combine the method with SSKK scheme. Both types of S&M relations are needed in proper test of the corrected and uncorrected data.

## 2. Dispersion relations for the powers of the complex reflectivity

The K-K relations are based on the causality principle [4, 5] and assumption that in the upper half of complex frequency plane (poles are located in the lower half plane) the complex reflection coefficient $r(\omega) = |r(\omega)|\exp\{i\varphi(\omega)\}$ (or complex refractive index $N(\omega)$) is a holomorphic [5, 6] function, which has a strong enough fall–off at high frequencies. The assumption of the function to be holomorphic in the upper half plane is valid if the incident electric field oscillates proportional to $\exp(-i\omega t)$. Since the function $r^n(\omega)$ is also holomorphic for any positive integer $n = 1, 2,\ldots$, it satisfies the following pair of S & M dispersion relations [3]

$$|r(\omega')|^n \cos[n\varphi(\omega')] = \frac{2}{\pi} P\int_0^\infty \frac{|r(\omega)|^n \sin[n\varphi(\omega)]}{\omega^2 - \omega'^2} d\omega$$

$$|r(\omega')|^n \sin[n\varphi(\omega')] = -\frac{2\omega'}{\pi} P\int_0^\infty \frac{|r(\omega)|^n \cos[n\varphi(\omega)]}{\omega^2 - \omega'^2} d\omega$$

(1)

where P stands for the Cauchy principal value. One can observe from (1) that in these relations, amplitude and phase are mixed up. However, the problem of data extrapolation beyond the measured spectral range is not as severe as that for the conventional K-K relations, which are presented in terms of $\log |r(\omega)|$ and $\varphi(\omega)$ [4, 5]. Fast fall-off of the power of the reflection coefficient at high frequency is a crucial property for strong convergence of the dispersion relations given in equation (1).

A singly subtractive K-K (SSKK) relation for the phase retrieval from the logarithm of reflectance was first employed by Ahrenkiel [6], and the concept was generalized by Palmer et al [7]. We apply this technique for the powers of the complex reflection coefficient $r^n(\omega)$, and obtain the relations as follows:

$$|r(\omega')|^n \cos[n\varphi(\omega')] = |r(\omega_1)|^n \cos[n\varphi(\omega_1)] + \frac{2(\omega'^2 - \omega_1^2)}{\pi} P\int_0^\infty \frac{\omega |r(\omega)|^n \sin[n\varphi(\omega)]}{(\omega^2 - \omega'^2)(\omega^2 - \omega_1^2)} d\omega$$

$$\frac{|r(\omega')|^n \sin[n\varphi(\omega')]}{\omega'} = \frac{|r(\omega_1)|^n \sin[n\varphi(\omega_1)]}{\omega_1} - \frac{2(\omega'^2 - \omega_1^2)}{\pi} P\int_0^\infty \frac{|r(\omega)|^n \cos[n\varphi(\omega)]}{(\omega^2 - \omega'^2)(\omega^2 - \omega_1^2)} d\omega$$

(2)

where $\omega_1$ is an anchor point where the reflection coefficient is known. In our calculations we have chosen the anchor point using the method suggested in Ref. 7. For the sake of simplicity we have chosen the same anchor point in both dispersion relations (2). Since in Eq. (2) the S&M and SSKK techniques have been combined together, we ill refer to Eq. (2) as the singly- subtractive Smith and Manogue (SS-S&M) dispersion relations. The merit of the SS- S&M dispersion relations is their high convergence at finite measured spectral range just like in the case of conventional S& M dispersion relations.

Both types of S&M- dispersion relations allow one to check consistency of the reflection data. Specifically, by using the measured amplitude and phase of the complex reflection coefficient one can calculate the corresponding real and imaginary parts from Eqs. (1) & (2) at frequency $\omega'$ that belongs to the frequency range in question. A large discrepancy between the calculated and measured complex reflectivity indicates a presence of an experimental error. This error may originate, for example, from the (negative) sample misplacement (spatial precision of a few micron) in the THz- setup[1, 2].

## 3. Testing the consistency of the THz reflection spectra

In the terahertz time-domain reflection spectroscopy, one can obtain simultaneously both the amplitude and the phase of the complex reflection coefficient from the time-domain waveform of the reflected electric field. These measurements are conventionally performed, by comparing the results obtained with the sample and a reference. In practice various experimental ambiguities distort the obtained complex spectra of the reflection coefficient. One of such ambiguities is the problem of sample position. Since it is virtually impossible to adjust the sample and the reference exactly at the same position, the phase of the THz-TDS carries a frequency dependent misplacement error. The amplitude of the signal can be considered to be correct because (i) measurements are performed in the far-field region and (ii) the reference reflectivity is frequency independent. Since, the frequency-dependent real and imaginary parts of the complex reflection coefficient are not correct, due to the phase error, the complex refractive index of opaque sample obtained in THz-TDS is not reliable. Fortunately, the phase error caused by sample misplacement can be corrected either by using maximum entropy model (MEM) [1] or SSKK method [2]. However, the consistency of the corrected data (i.e. whether real and imaginary parts of the reflection coefficient fulfill the causality principle) should be checked using dispersion theory.

Next we apply two the S &M and SS- S&M dispersion relations for the measured reflection spectra of *n*-type, undoped (100) InAs wafer in the spectral range 0.5 – 2.5 THz. The data was obtained with oblique incidence about 45 degrees for *p*-polarized THz radiation. Because of the large beam diameter of the THz wave, the incidence angle was determined by fitting which returned the value of 35 degrees. The procedure will be described later. As a reference we used aluminum plate and it is assumed that its reflection coefficient is unity in the THz spectral range of interest. The experimental setup and detailed description of the sample were reported in Ref. 8.

From the measured amplitude and the phase spectra shown in Fig. 1 one can observe that there is a phase error due to negative sample displacement. In THz spectral region, the permittivity can be obtained in terms of the classical Drude model that yields

$$\varepsilon(\omega) = \varepsilon_b (1 + \frac{A}{-\omega^2 - i\Gamma\omega}), \quad (3)$$

where $\varepsilon_b$ and $\Gamma$ are the background permittivity and damping factor, respectively. In Fig. 1 we show also the amplitude and phase calculated with Drude model and Fresnel's formula for reflection of *p*-polarized radiation at oblique light incidence. One can observe in Fig. 1 that the measured and the calculated amplitude match pretty well, while the experimental and the calculated phase of the reflection coefficient do not correspond to each other. The reason is the sample misplacement error, which can be removed by either methods presented in [1, 2]. The phase calculated by these two methods matches well with that obtained using the Drude model for permittivity [1, 2].

In Fig. 2 we show the erroneous real and imaginary parts of the complex reflection coefficient calculated using uncorrected data presented in Fig. 1 along with those obtained from the S&M and SS- S&M analyses for the case *n = 1*. One can observe that the calculated curves depart to a great extent from the experimental ones. Hence we may draw a tentative conclusion that the measured THz data is not correct. It is necessary to emphasize that we have utilized *pairs of dispersion relations*, whereas in conventional K-K analysis only the other partner of the K-K relations is exploited.

It should be noted that the algorithms based on S&M dispersion relations are applied for the real and imaginary parts of the complex reflection coefficient, instead of the measured quantities, which are the amplitude and the phase. This brings some limitation in the analysis.

For the validity of the conventional K-K- and both types of S&M dispersion relations the true complex reflection coefficient has to be a holomorphic function in the upper half of complex plane, $\varpi = \omega + i\upsilon$, $\upsilon > 0$. It must also have sufficient decay for high complex frequency and fulfill a symmetry relation due to the parity of the complex reflection coefficient. Specifically, the phase error due to sample displacement can be expressed as

$$\Delta\varphi = \frac{\Delta L}{c}\omega, \qquad (4)$$

where *ΔL* is misplacement error (difference in the optical path between the sample and reference) and *c* is the speed of light in the vacuum. Since the spot size of the incident THz beam on the sample surface is about 0.3mm in the case of present data, hence the ratio *ΔL/c* is constant. One can expect that the real and imaginary parts of the reflection coefficient are subject to an error when the phase error is large enough as in the case of Fig. 2. The erroneous reflection coefficient can be given as follows:

$$r_{err}(\omega) = |r_{err}(\omega)| e^{i(\varphi(\omega) + \frac{\Delta L}{c}\omega)}. \tag{5}$$

At $\Delta L > 0$ this function is holomorphic in the upper half plane and it fulfills the symmetry relations imposed on the true reflection coefficient. However, one can readily find that conventional K-K analysis, which is based on the logarithm of the complex reflection coefficient, becomes invalid because $Im\{ln\ |r_{err}|\} \to \infty$ as the frequency tends to infinity. At the same time, $r_{err}$, satisfies S&M and SS- S&M dispersion relations for a *positive* misplacement error $\Delta L \geq 0$, i. e. this error can not be identified by the S&M dispersion analysis. The situation is totally different for the case of negative sample misplacement. Indeed, when we extent the exponent function involving the incorrect phase in Eq. (5), into the upper half plane, $\varpi = \omega + i\upsilon$, we observe that we have to break the property of the reflection coefficient to tend to zero as the modulus of the complex frequency tends to infinity. Thus the function $exp(|\Delta L|\upsilon/c)$ blows up as $\upsilon$ tends to infinity. Hence, the validity of K-K, S&M and SS- S&M dispersion relations is broken, since the asymptotic fall-off of the reflection coefficient in the upper half plane is no more valid. This is because the Jordan's lemma needed in the complex contour integration of derivation of K-K or modified K-K relations is no more valid (see Appendix C in Ref. 4). There is no break up of causality, merely it is the mathematical property related to the definition of complex reflection coefficient by an exponential function. Breaking of the validity of the S&M and SS-S&M dispersion relations manifest itself as discrepancy between inverted and experimental data**.** Limitation of the sign of misplacement is resulting from the procedure of the analyses, where we convert the phase and the amplitude data to real and imaginary part prior to the analysis.

It is interesting that the validity of the K-K- and two types of S&M dispersion relations depends on the experimental set up. As we already discussed this has purely mathematical origin, and there is no break up of the principle of causality. In THz experiments one can intentionally make misplacement error to be negative so that the error can be identified by the two types of S&M dispersion relations.

Naturally when we find the phase error and correct the data according to it, we find agreement between the measured and inverted data. One should apply the two types of S&M dispersion relations at two stages, one is at the point of erroneous real and imaginary parts, and the other is after correction of misplacement. This means that we can test how good estimate we have obtained for the complex reflection coefficient after correction of the misplacement error of the sample in the relevant spectra.

Real part of the reflection coefficient of InAs calculated by the two types of S&M dispersion relations for $n = 1$, using the corrected experimental data, is presented in Fig. 3. It is interesting to observe from Fig. 3 that the SS-S&M analysis reproduces the real part from the imaginary part pretty well, whereas in the curve obtained with S& M there is an offset. However, it is important to calculate the real part using both S & M dispersion relations, as it will be described below. An approximation of this offset can be obtained by averaging the difference of the two curves over the relevant spectral range. It can be shown that if we lift the S&M curve in Fig. 3 by the offset value, a little bit better approximation for $Re\{r(\omega)\}$ is obtained than in the case of SS-S&M. However, if we just substitute the obtained correct real part of the reflection coefficient into relevant dispersion relation we get absurd values for $Im\{r(\omega)\}$ with both types of S&M analyses. The reason is that in order to calculate the imaginary part of the reflection coefficient we have to take into account the finite spectral interval of the experimental data. Specifically, the derivation of the dispersion relations involves an identity

$$P\int_0^\infty \frac{C}{\omega^2 - \omega'^2} d\omega = 0, \qquad (6)$$

where $C$ is a constant. In the case of a finite spectral range, the integration is determined by the frequency interval of the experiment and, correspondingly, identity (6) can no longer be employed, and right hand side of Eq. (6= should be replaced with a finite value. It can be shown that in such a case we arrive at finite, frequency-independent offset, $r_{offset}$, to the real part of the reflection coefficient in the S&M analyses. This offset has to be taken into account in the analysis of the experimental spectra using dispersion relations. Thus in order to calculate the imaginary part of the reflection coefficient one needs to replace $r$ with $r$- $r_{offset}$ in S&M and SS-S&M dispersion relations. In order to obtain $r_{offset}$ we need first to employ the S&M dispersion relation for $n = 1$. After the subtraction of the offset from the real part of reflection coefficient the imaginary part is calculated for high power $n$. Naturally high power technique can be applied also to test the real part. In Fig. 4 we show the curves for $n = 10$ for $\text{Re}\{(r-r_{offset})^{10}\}$ and $\text{Im}\{(r-r_{offset})^{10}\}$, which were calculated using both types of S&M dispersion relations and the Drude model for fitting the experimental data. The curves match very well with the exact ones. Obviously high power of reflection coefficient is effective in testing band-limited data. One can freely select which to use of the two types of S&M or (or both) SS-S&M dispersion relations for high power $n$. The real and imaginary parts of the complex reflection coefficient can be calculated by taking $10^{th}$ root of the data of Fig. 4.

## 5. Conclusions

In conclusion, we have demonstrated that S&M and SS-S&M dispersion relations for the integer power of complex reflectivity provide us with powerful tools in order to improve data inversion at a relatively narrow spectral range. We compared the S&M and SS-S&M analyses and observed that in the present cases S&M dispersion relations perform as good as SS-S&M relations, which need phase information at one anchor point. As an application the developed analysis can be applied to verify the consistency of the experimental data in terahertz spectroscopy, for the case of negative misplacement, when both amplitude and phase of the reflection coefficient can be measured, and the success of phase correction. The rigorous treatment of positive misplacement error and the extension of the algorithm for application to the other type of systematic error such as diffraction effects are open for future study.

**Figure captions**

Fig. 1 Upper panel: measured amplitude of In As at terahertz range (filled circles) and amplitude obtained from Drude model (solid line). Lower panel: measured phase (filled circles), phase obtained from Drude model (solid line) and corrected phase obtained by the methods in Refs. 1 & 2 (filled triangles). (b) incorrect real and imaginary parts of the reflection coefficient (solid lines).

Fig. 2. Incorrect real and imaginary parts of the reflection coefficient (solid lines). The curves obtained by S & M (dashed) and with SS- S&M (dotted line) analyses were calculated using the measured data (solid lines) for the case $n = 1$.

Fig. 3 Real part of the reflection coefficient of InAs calculated from S&M (dashed line) and from SS-S&M (dotted line) dispersion relations for the case $n = 1$. The solid line presents the real part, which was obtained after phase correction by methods of Refs. 1 & 2.

Fig. 4. Re$\{(r-r_{offset})^{10}\}$, and b) Im$\{(r-r_{offset})^{10}\}$ of InAs calculated from S&M (dotted line) and from SS-S&M (dashed line) dispersion relations in the case $n = 10$. The solid lines present the curves obtained using Drude model for InAs. The filled circles correspond to experimental data points.

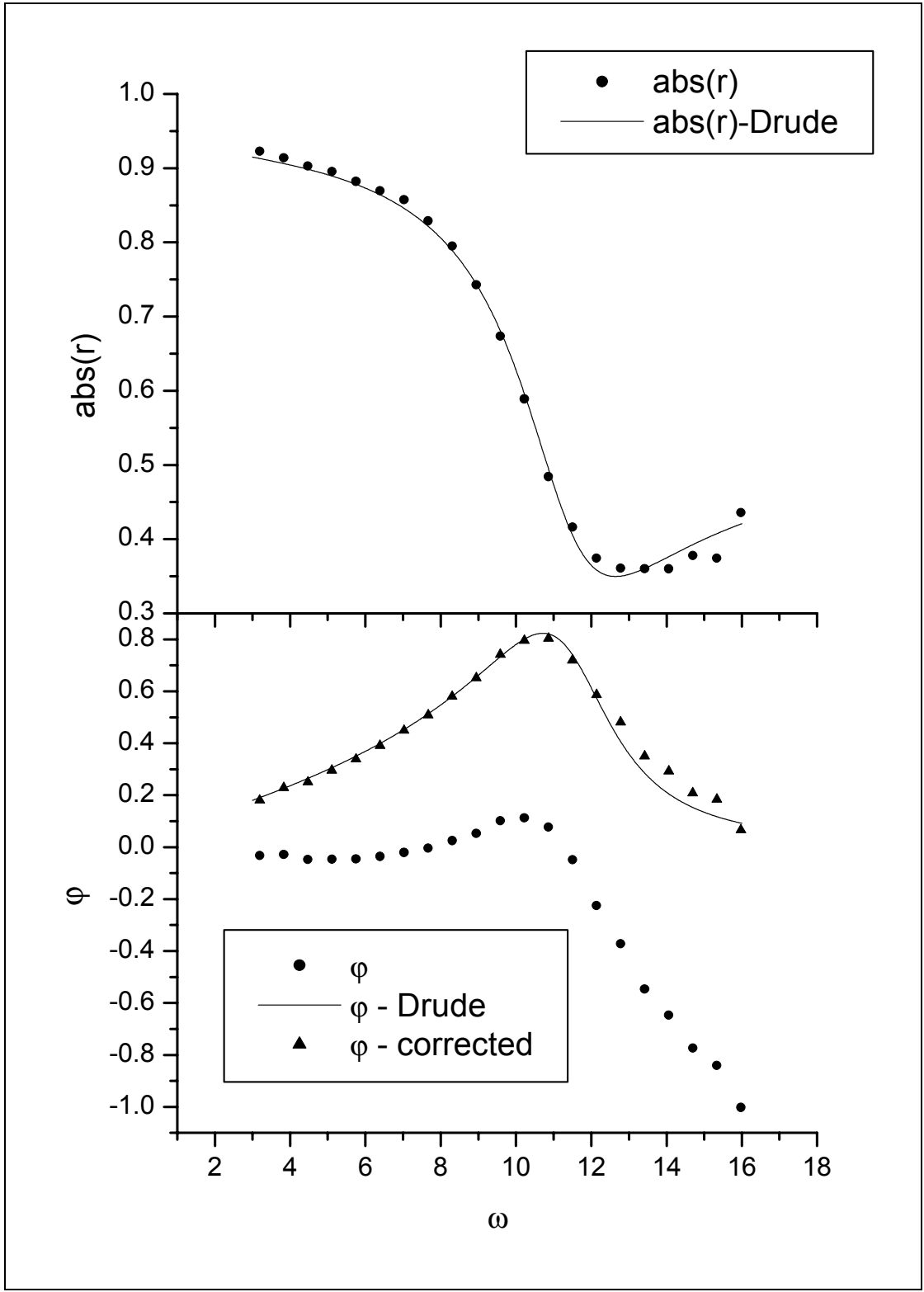

Fig.1 Peiponen et al

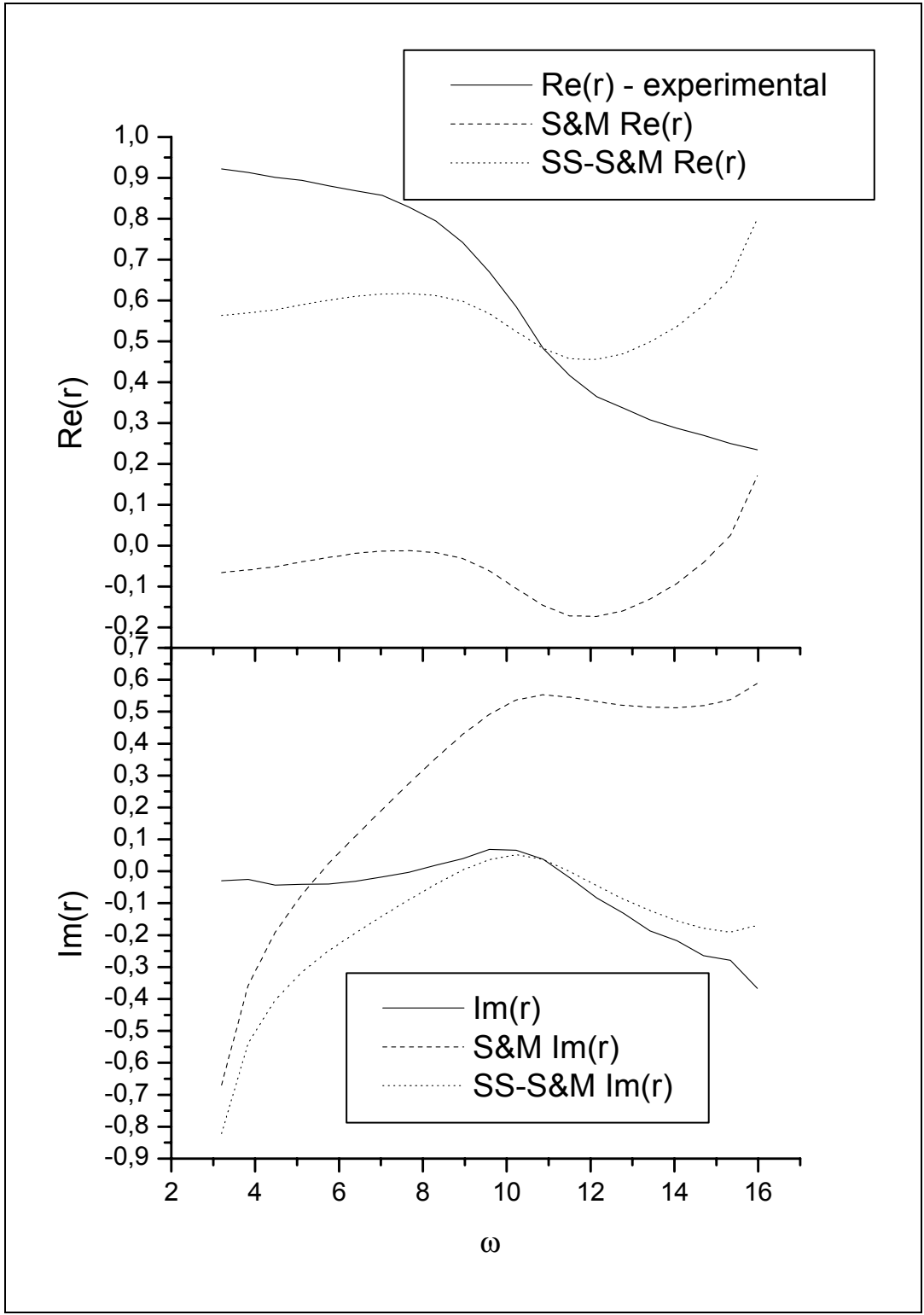

Fig. 2 Peiponen et al

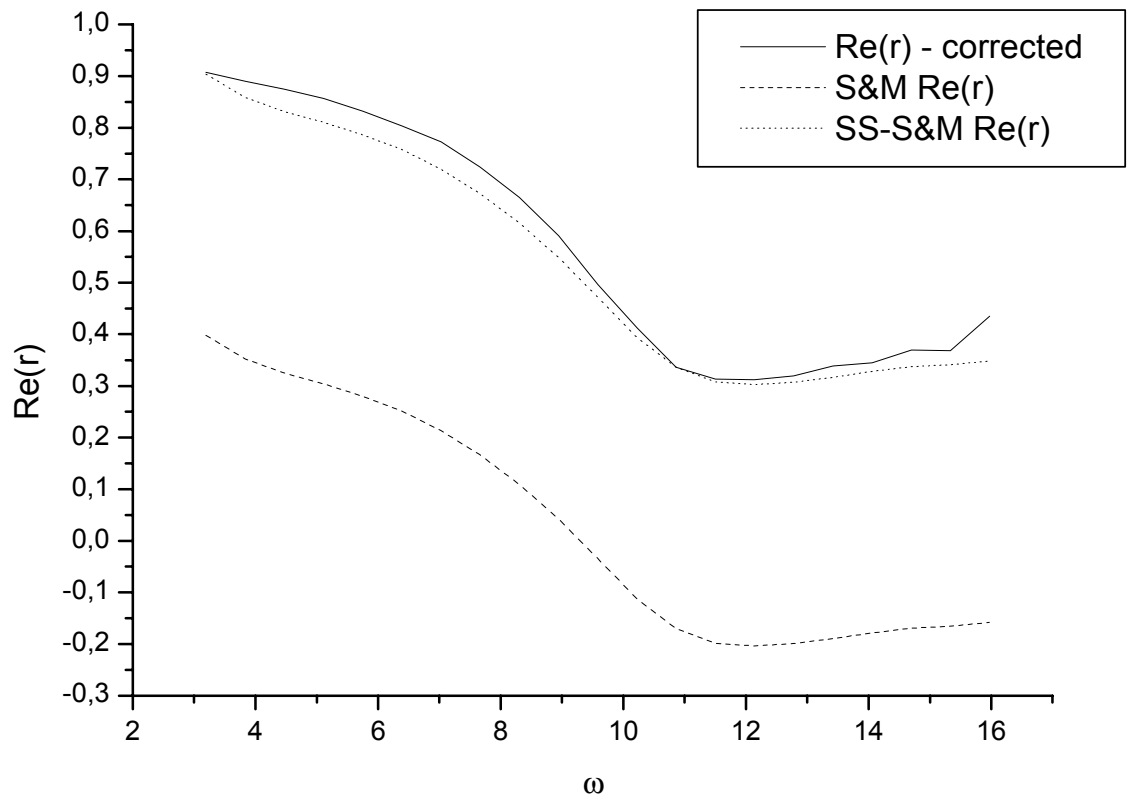

Fig. 3 Peiponen et al

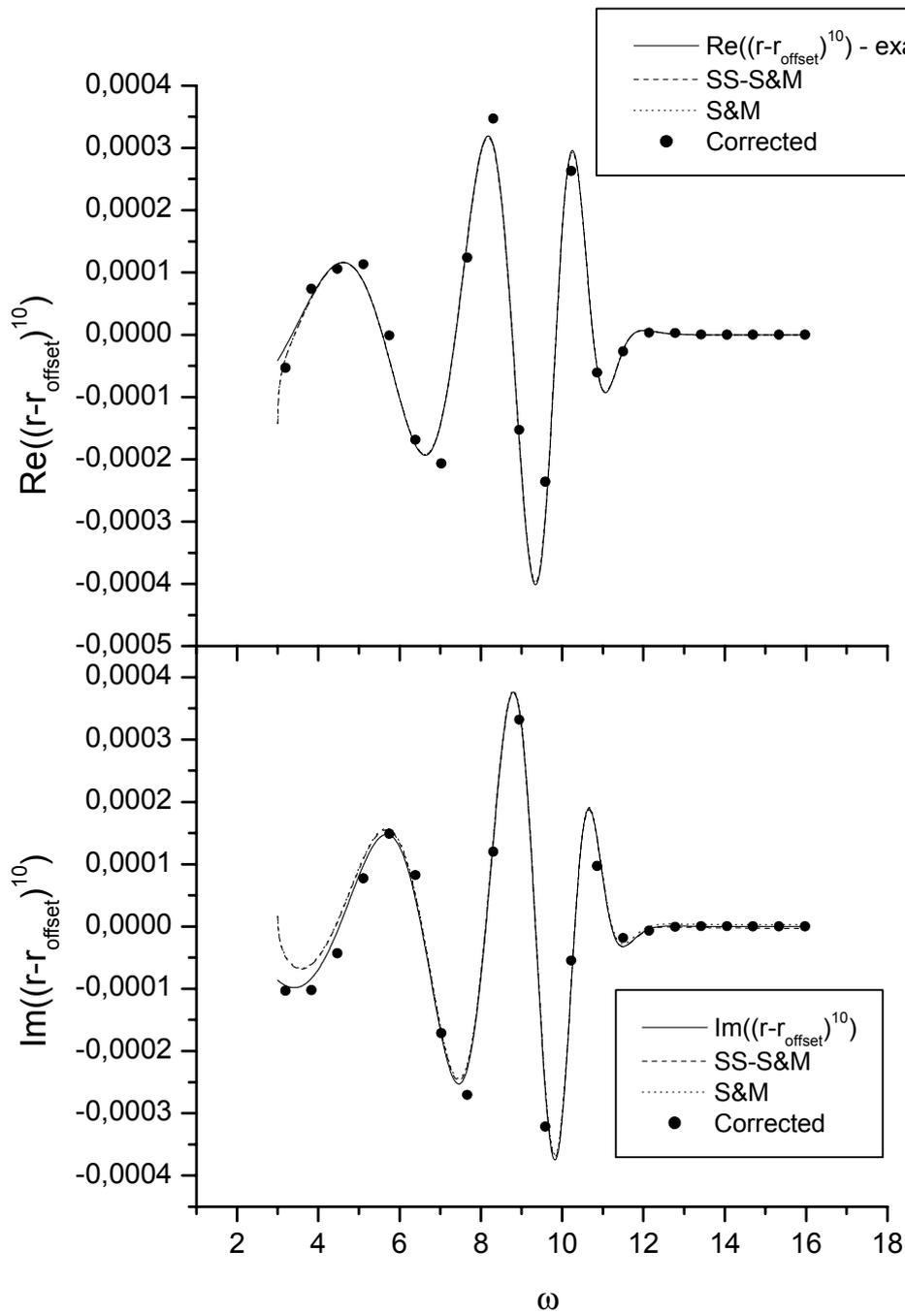

Fig. 4 Peiponen et al